\documentclass[prd,nofootinbib]{revtex4}
\usepackage{epsfig,amsmath,amssymb,slashed}
\usepackage[colorlinks=true,linktocpage=true,linkcolor=blue,citecolor=blue]{hyperref}

\newcommand{\di}{{\rm d}}
\newcommand{\ii}{i}

\def\spt{{\cal S}}
\def\wT{{\widehat T}}
\def\wj{{\widehat j}}
\def\wspt{{\widehat{\cal S}}}

\def\wPhi{{\widehat{\Phi}}}

\def\wpsi{{\widehat{\psi}}}
\def\wrho{{\widehat{\rho}}}

\def\wP{{\widehat{P}}}
\def\wJ{{\widehat{J}}}
\def\wQ{{\widehat{Q}}}

\newcommand{\tr}{{\rm tr}}  
  
\newcommand{\e}{{\rm e}}

\newcommand{\be}{\begin{equation}}
\newcommand{\ee}{\end{equation}}                                                                               
\newcommand{\bea}{\begin{eqnarray}}
\newcommand{\eea}{\end{eqnarray}}

\begin{document}

\title{Spin tensor and its role in non-equilibrium thermodynamics}

\author{Francesco Becattini}
\affiliation{Universit\`a di Firenze and INFN Sezione di Firenze, Via G. Sansone 1, I-50019, Sesto Fiorentino (Firenze), Italy} 
\author{Wojciech Florkowski}
\affiliation{Institute of Nuclear Physics Polish Academy of Sciences, PL-31342 Krak\'ow, Poland}
\author{Enrico Speranza}
\affiliation{Institute for Theoretical Physics, Goethe University, D-60438 Frankfurt am Main, Germany}

\begin{abstract}
It is shown that the description of a relativistic fluid at local thermodynamic equilibrium 
depends on the particular quantum stress-energy tensor operator chosen, e.g., the canonical 
or symmetrized Belinfante stress-energy tensor. We argue that the Belinfante tensor is not 
appropriate to describe a relativistic fluid whose macroscopic polarization relaxes slowly 
to thermodynamic equilibrium and that a spin tensor, like the canonical spin tensor, is 
required. As a consequence, the description of a polarized relativistic fluid involves an 
extension of relativistic hydrodynamics including a new antisymmetric rank-two tensor as a 
dynamical field. We show that the canonical and Belinfante tensors lead to different 
predictions for measurable quantities such as spectrum and polarization of particles produced 
in relativistic heavy-ion collisions.
\end{abstract}

\maketitle

\section{Introduction}
\label{intro}

The measurement of a finite global polarization of particles in relativistic heavy-ion 
collisions \cite{STAR}, in agreement with the predictions of relativistic hydrodynamics 
\cite{becacsernai,becavort} (see also \cite{Pang:2016igs,Deng:2016gyh,Karpenko:2016jyx,
Xie:2017upb,Li:2017slc,Sun:2017xhx,Lan:2017nye,Xia:2018tes}), has opened a new perspective 
in the phenomenology of these collisions as well as in the theory of relativistic 
matter, showing for the first time a direct manifestation of quantum features in 
this realm. While a formula relating mean polarization with thermal vorticity (see 
Sec.~\ref{leqdo}) at local thermodynamic equilibrium was obtained in Ref.~\cite{becaspin}, 
based on an educated {\em ansatz}, an exact formula is still missing even at global 
thermodynamic equilibrium with rotation. Meanwhile, the experiments have proved to 
be able to probe polarization differentially in momentum space \cite{STAR2,QMstar} 
and, from the theory standpoint, the issue has been raised \cite{flork} about the 
relevance of the spin tensor in the description of a relativistic fluid. 

Indeed, the problem of the physical significance of the spin tensor --- mostly in relativistic 
gravitational theories, notably in the Einstein-Cartan theory --- is a long-standing one \cite{hehl} 
and has been rediscussed more recently in Refs.~\cite{becatinti1,becatinti2}, where it was demonstrated 
that well known thermodynamic formulae for transport coefficients such as viscosity do depend on 
the presence of a spin tensor. However, up to now, its relevance for the polarization measured 
in relativistic heavy-ion collisions has not been discussed in detail, even though it was used 
to derive the polarization formula in Ref.~\cite{becaspin}. It is one of the main 
goals of this paper to present and fully explore this theoretical issue in detail. 

Our conclusion is that for a general relativistic fluid the spin tensor is significant, if 
spin density ``slowly" relaxes towards equilibrium, where slowly means on a time scale which 
is comparable to the familiar hydrodynamic time scale of evolution of charge and momentum 
densities. In this case, we will see that the description of the polarization of particles and 
the calculation of its final value requires a new thermodynamic potential,  akin to chemical 
potential or temperature, coupled to the spin tensor. Accordingly, relativistic hydrodynamics 
should be extended to include the spin tensor among the evolved densities.

The paper is organized as follows: In Sec.~\ref{pseudo} we define the spin tensor 
and its close relationship with the stress-energy tensor. Section~\ref{leqdo} describes 
the local thermodynamic equilibrium density operator and its dependence on the 
pseudo-gauge transformations of the stress-energy and spin tensors. In Sec.~\ref{ineq} 
we discuss the physics of different local thermodynamic equilibria, while in Sec.~\ref{measur} 
we consider consequences of using a spin tensor for measurable quantities such as 
spectra and polarization. We summarize and conclude in Sec.~\ref{discu}.
  

\medskip
{\it Notation:} In this paper we adopt the natural units, with $\hbar=c=k_{\rm B}=1$. 
The Minkowski metric tensor is ${\rm diag}(1,-1,-1,-1)$; for the Levi-Civita symbol 
we use the convention $\varepsilon^{0123}=1$. We use the relativistic notation with
repeated indices assumed to be saturated. Operators in Hilbert space are denoted 
by an upper hat, e.g., $\widehat A$. Although we work in flat space-time, in many 
equations we use covariant derivative $\nabla_\mu$ instead of partial derivative 
$\partial_\mu$ to emphasize the validity of the relations in general coordinates.

\section{Pseudo-gauge transformations}
\label{pseudo}

In relativistic quantum field theory in flat space-time, according to Noether's theorem, for 
each continuous symmetry of the action there is a corresponding conserved current. 
The currents associated with the translational symmetry and the Lorentz symmetry
~\footnote{By Lorentz symmetry transformations we understand here Lorentz boosts 
and rotations.} are the so-called {\em canonical} stress-energy tensor and the {\em canonical} 
angular momentum tensor:
\begin{align}\label{cantens}
  \wT^{\mu\nu}_C &= \sum_a \frac{\partial \cal L}{\partial(\partial_\mu \wpsi^a)} \partial^\nu \wpsi^a
   - g^{\mu\nu} {\cal L}, \\
  \widehat{\cal J}^{\mu,\lambda\nu}_C &= x^\lambda \wT^{\mu\nu}_C - x^\nu \wT^{\mu\lambda}_C 
   + \wspt^{\mu,\lambda\nu}_C.
\end{align}
Here ${\cal L}$ is the lagrangian density, while $\wspt_C$ reads
\begin{eqnarray} \label{SC}
   \wspt^{\mu,\lambda\nu}_C =  - \ii \sum_{a,b} \frac{\partial \cal L}{\partial(\partial_\mu \wpsi^a)} 
    D(J^{\lambda\nu})^a_b \wpsi^b
\end{eqnarray}  
with $D$ being the irreducible representation matrix of the Lorentz group pertaining 
to the field. The above tensors fulfill the following equations:
\be\label{conserv}
\nabla_\mu \wT^{\mu\nu}_C = 0, 
\qquad \nabla_\mu  \widehat{\cal J}^{\mu,\lambda\nu}_C
= \wT^{\lambda\nu}_C - \wT^{\nu\lambda}_C + \nabla_\mu \wspt^{\mu,\lambda\nu}_C = 0,
\ee 

It turns out, however, that the stress-energy and angular momentum tensors are not 
uniquely defined. Different pairs can be generated either by just changing the lagrangian 
density or, more generally, by means of the so-called pseudo-gauge transformations \cite{hehl}:
\begin{eqnarray}\label{transfq}
 && \wT^{\prime \mu \nu} = \wT^{\mu \nu} +\frac{1}{2} \nabla_\lambda
 \left( \wPhi^{\lambda, \mu \nu } - \wPhi^{\mu, \lambda \nu} - 
 \wPhi^{\nu, \lambda \mu}  \right), \nonumber \\
 && \wspt^{\prime \lambda, \mu \nu} = \wspt^{\lambda,\mu\nu}-\wPhi^{\lambda,\mu\nu},
\end{eqnarray}
where $\wPhi$ is a rank-three tensor field antisymmetric in the last two indices
(often called and henceforth referred to as {\em superpotential}). In Minkowski 
space-time, the newly defined tensors preserve the total energy, momentum, and angular 
momentum (herein expressed in Cartesian coordinates):
\begin{eqnarray}\label{total}
  \wP^\nu = \int_\Sigma \di \Sigma_\mu \wT^{\mu\nu}, \qquad 
 \wJ^{\lambda\nu} = \int_\Sigma \di \Sigma_\mu \widehat{\cal J}^{\mu,\lambda\nu},
\end{eqnarray}
as well as the conservation equations (\ref{conserv}) \footnote{This statement only
applies to Minkowski space-time, in generally curved space-times it is no longer
true \cite{hehl}.}.

A special pseudo-gauge transformation is the one where one starts with the canonical 
definitions and the superpotential is the spin tensor itself, that is, $\wPhi = \wspt$. 
In this case, the new spin tensor vanishes, $\wspt^\prime = 0$, and the new stress-energy tensor 
is the so-called Belinfante stress-energy tensor $\wT_B$,
\be\label{belinf}
 \wT^{\mu \nu}_B = \wT^{\mu \nu}_C +\frac{1}{2} \nabla_\lambda
 \left( \wspt^{\lambda, \mu \nu}_C - \wspt^{\mu, \lambda \nu}_C - 
 \wspt^{\nu, \lambda \mu}_C  \right).
\ee
%

\section{Local equilibrium density operator}
\label{leqdo}

The density operator describing local thermodynamic equilibrium in quantum field 
theory was obtained in ref.~\cite{zubarev,weert} and has been rederived more recently 
in refs.~\cite{betaframe,hongo}; herein, we briefly summarize the derivation.
The local thermodynamic equilibrium density operator is obtained by maximizing the 
entropy $S= -\tr (\wrho \log \wrho)$ with the constraints of given mean densities of 
conserved currents over some space-like hyper-surface $\Sigma$, a~covariant generalization 
of a hyperplane in special relativity. The projections of the mean stress-energy tensor 
and charge current onto the normalized vector perpendicular to $\Sigma$ must be 
equal to the actual ones:
\be\label{constr}
n_\mu \tr \left(\wrho \, \wT^{\mu\nu}\right) = n_\mu T^{\mu\nu}, 
\qquad n_\mu \tr \left(\wrho \, \wj^{\mu}\right) = n_\mu j^{\mu}.
\ee
where the operators are in the Heisenberg representation.
In addition to the energy, momentum, and charge densities, one should include the angular 
momentum density amongst the constraints in Eq.~(\ref{constr}), namely:
\be\label{constr2}
   n_\mu \tr \left( \wrho \, \widehat{\cal J}^{\mu,\lambda\nu} \right) 
   = n_\mu \tr \left[ \wrho \left( x^\lambda \wT^{\mu\nu} - 
    x^\nu \wT^{\mu\lambda} + \wspt^{\mu,\lambda\nu} \right) \right] 
    = n_\mu {\cal J}^{\mu,\lambda\nu}.
\ee
However, it is clear that if we have Belinfante's stress-energy tensor $\wT_B$ 
with associated vanishing spin tensor, equation (\ref{constr2}) is redundant, since 
the angular momentum density constraint is implied in (\ref{constr}). Hence, 
Eq.~(\ref{constr}) remains the only relevant condition. The resulting operator
reads (the subscript LE stands for Local Equilibrium):
\be\label{leqd}
\wrho_{\rm LE} = \frac{1}{Z} \exp \left[-\int_\Sigma \di \Sigma_\mu \left(\wT^{\mu\nu}_B \beta_\nu 
 - \zeta \, \wj^\mu \right) \right],
\ee
where $\beta$ and $\zeta$ are the relevant Lagrange multiplier functions for this 
problem, whose meaning is the four-temperature vector and the ratio between local 
chemical potential and temperature, respectively \cite{betaframe}.

We note that the operator (\ref{leqd}) is not the actual density operator, because 
it is not generally stationary as required in the Heisenberg picture. In fact, the 
true density operator for a system in local thermodynamic equilibrium coincides 
with that given by Eq.~(\ref{leqd}) {\em at the initial time} $\tau_0$, that is, 
with $\Sigma \equiv \Sigma(\tau_0)$,
\be\label{truerho}
\wrho = \frac{1}{Z} \exp \left[ -\int_{\Sigma(\tau_0)} \di \Sigma_\mu \left(\wT^{\mu\nu}_B \beta_\nu 
 - \zeta \wj^\mu \right) \right].
\ee
Provided that fluxes at some timelike boundary vanish, it is possible to rewrite 
the actual density operator $\wrho$ by using Gauss' theorem \cite{betaframe},
\be\label{dynop}
\wrho = \frac{1}{Z} \exp \left[-\int_{\Sigma(\tau)} \di \Sigma_\mu \left(\wT^{\mu\nu}_B \beta_\nu - \zeta 
 \wj^\mu \right) + \int_\Theta \di \Theta \; \left( \wT^{\mu\nu}_B \nabla_\mu \beta_\nu - 
 \wj^\mu \nabla_\mu \zeta \right) \right],
\ee
where the first term is the local thermodynamic equilibrium term at time $\tau$ 
and $\Theta$ denotes the space-time region encompassed by the space-like hypersurfaces 
$\Sigma(\tau)$, $\Sigma(\tau_0)$, and the time-like boundaries. Formula (\ref{dynop}) 
is the essence of Zubarev's formalism \cite{zubarev,weert,becacov} and it nicely 
separates the non-dissipative part (the first term in the exponent) from the dissipative 
one (the second term)~\footnote{We note in passing that equation (\ref{dynop}) 
is basically the one used to obtain the first derivation of the Kubo formula of 
the shear viscosity~\cite{hosoya}.}.

The operator (\ref{dynop}) becomes stationary, that is independent of the hypersurface $\Sigma$, 
when the four-temperature is a Killing vector field and the ratio between chemical potential 
and temperature is constant:
\be\label{kill}
 \nabla_\mu \beta_\nu + \nabla_\nu \beta_\mu = 0, \qquad \nabla_\mu \zeta = 0.
\ee
The first equation above follows from the fact that Belinfante's energy-momentum tensor is symmetric. In a flat spacetime, it implies that the second order gradients of $\beta$ vanish and
$\beta$ itself is given by the expression
\be\label{killsol}
  \beta_\nu = b_\nu + \varpi_{\nu\lambda} x^\lambda,
\ee  
where $b$ is a constant vector and $\varpi$ is an antisymmetric tensor with constant 
coefficients. The conditions (\ref{kill}) define global thermodynamic equilibrium. 
One can also check that the (redundant) inclusion of the conservation of angular 
momentum (\ref{constr2}) in Belinfante's case (where it is reduced to the conservation 
of the orbital part only) does not change the form of global equilibrium, as this 
leads to a change of the tensor $\varpi_{\nu\lambda}$ which remains constant.

One can now use the pseudo-gauge transformations of Sec.~\ref{pseudo} to rewrite
the local thermodynamic equilibrium density operator as a function of, e.g., canonical 
tensors. Using Eq.~(\ref{belinf}), we find:
\begin{eqnarray}
\label{leqdC}
\wrho_{\rm LE} 
&=& \frac{1}{Z} \exp \left[-\int_\Sigma \di \Sigma_\mu \left( \wT^{\mu\nu}_B 
\beta_\nu  - \zeta \wj^\mu \right) \right] \nonumber \\
&=& \frac{1}{Z} \exp \left[-\int_\Sigma \di \Sigma_\mu \left( \wT^{\mu\nu}_C 
  \beta_\nu + \frac{1}{2} \beta_\nu \nabla_\lambda \left( \wspt^{\lambda, \mu \nu}_C 
- \wspt^{\mu, \lambda \nu}_C - \wspt^{\nu, \lambda \mu}_C  \right)- \zeta \wj^\mu \right) \right].
\end{eqnarray}
We now work out the integral involving the spin tensor. Integrating by parts we obtain
\be\label{int1}
 -\frac{1}{2} \int_\Sigma \di \Sigma_\mu 
 \left[ \; \nabla_\lambda \left( \beta_\nu 
 \wspt^{\lambda, \mu \nu}_C - \beta_\nu \wspt^{\mu, \lambda \nu}_C - \beta_\nu 
 \wspt^{\nu, \lambda \mu}_C \right) \right]
 + \frac{1}{2} \int_\Sigma \di \Sigma_\mu \left[
\nabla_\lambda \beta_\nu \left( \wspt^{\lambda, \mu \nu}_C - \wspt^{\mu, \lambda \nu}_C 
  - \wspt^{\nu, \lambda \mu}_C \right) \right] .
\ee
The first term is a divergence of an antisymmetric tensor (with respect to the 
$\lambda \leftrightarrow \mu$ exchange), so it can be turned into a boundary integral,
\be\label{term1}
 -\frac{1}{4}\int_{\partial \Sigma} \di \tilde S_{\mu\lambda} \; \left( \beta_\nu 
 \wspt^{\lambda, \mu \nu}_C - \beta_\nu \wspt^{\mu, \lambda \nu}_C - 
 \beta_\nu \wspt^{\nu, \lambda \mu}_C \right),
\ee
which vanishes for suitable boundary conditions imposed on $\beta$ and/or $\wspt$. 
The second term, on the other hand, can be rewritten as
\be\label{int2}
 -\frac{1}{2} \int_\Sigma \di \Sigma_\mu \; \left[
 \nabla_\lambda \beta_\nu \wspt^{\mu, \lambda \nu}_C
  - \nabla_\lambda \beta_\nu
  \left ( \wspt^{\lambda, \mu \nu}_C + \wspt^{\nu, \mu \lambda}_C \right)  \right]
\ee
by taking advantage of the antisymmetry of the last two indices. In this way, we 
finally obtain
\be\label{leqdC2}
\wrho_{\rm LE} = \frac{1}{Z} \exp \left[-\int_\Sigma \di \Sigma_\mu \left(\wT^{\mu\nu}_C 
  \beta_\nu - \frac{1}{2} \varpi_{\lambda\nu} \wspt^{\mu, \lambda \nu}_C - \frac{1}{2}
  \xi_{\lambda\nu} \left( \wspt^{\lambda, \mu \nu}_C + \wspt^{\nu, \mu \lambda}_C \right)  
   - \zeta \wj^\mu\right) \right],
\ee
where
\be\label{thvort}
\varpi_{\lambda\nu} = \frac{1}{2} (\nabla_\nu \beta_\lambda - \nabla_\lambda \beta_\nu)
\ee
is the thermal vorticity, compare Eq.~(\ref{killsol}), and 
\be\label{symm}
\xi_{\lambda\nu} = \frac{1}{2} (\nabla_\nu \beta_\lambda + \nabla_\lambda \beta_\nu)  
\ee
is the symmetric part of the gradient of the four-temperature vector
\footnote{
It is worth noting that for the free Dirac field, the canonical spin tensor has the form
\be\label{DiracS}
  \wspt^{\mu,\lambda\nu}_C = \frac{i}{8}\, 
  {\overline {\widehat \psi}} \{\gamma^\mu , [\gamma^\lambda,\gamma^\nu] \} {\widehat \psi},
\ee
which is completely antisymmetric, the term proportional to $\xi$ in Eq.~(\ref{leqdC2}) vanishes.
It has been recently pointed out that a non-completely antisymmetric spin tensor can be connected
to a non-vanishing energy dipole moment \cite{lorce}}. 

The conclusion we can draw from Eq.~(\ref{leqdC2}) is apparent. If we had used 
the canonical stress-energy tensor instead of Belinfante's form, enforcing only 
the constraints (\ref{constr})  with $\wT_C$ replacing $\wT_B$ to maximize the 
entropy, we would have obtained a formally analogous expression for the local 
thermodynamic equilibrium density operator, i.e., Eq.~(\ref{leqd}) with $\wT_B$ 
replaced by $\wT_C$. However, as shown above, the new local thermodynamic equilibrium 
operator would have not been the same as that given by Eq.~(\ref{leqdC2}). Therefore, 
in general, there is no equivalence of description of local thermodynamic equilibrium 
with different sets of tensors. This conclusion is not surprising after all because 
densities rather than integrated quantities are fixed by the constraints in local 
thermodynamic equilibrium and densities do depend on the pseudo-gauge choice (\ref{transfq}): 
the concept of local thermodynamic equilibrium is not pseudo-gauge independent. 

One may wonder, however, if the use of canonical tensors simply requires the inclusion 
of Eq.~(\ref{constr2}) besides Eq.~(\ref{constr}). In fact, since Eq.~(\ref{leqd}) 
is equivalent to Eq.~(\ref{leqdC2}), by inclusion of the spin tensor within the 
canonical approach, one can indeed obtain the same $\wrho_{\rm LE}$ for the canonical 
and Belinfante's schemes. This issue will be discussed in greater detail in the next 
section. Nevertheless, at global thermodynamic equilibrium, the equivalence is fully 
restored: the tensor (\ref{symm}) vanishes according to (\ref{kill}), the four-temperature 
vector is given by (\ref{killsol}), $\varpi$ is constant and the density operator becomes:
\be\label{intrho}
 \wrho =  \frac{1}{Z} \exp \left[- b_\mu \wP^\mu + \frac{1}{2} \varpi_{\mu\nu} \wJ^{\mu\nu} + \zeta \wQ \right],
\ee
where $\wP^\mu$ and $\wJ^{\mu\nu}$ are given by Eq.~\eqref{total} and $\wQ$ is the 
total charge defined as
\be
\wQ = \int_\Sigma \di \Sigma_\mu \wj^\mu .
\ee
The form \eqref{intrho}, depending on the generators, is manifestly invariant under 
any pseudo-gauge transformation.

\section{Local thermodynamic equilibrium with spin tensor and spin hydrodynamics}
\label{ineq}

As it has been mentioned at the end of the previous section, it seems compelling to 
include angular momentum density amongst the constraints defining local thermodynamic 
equilibrium, see Eq.~(\ref{constr2}), in case one starts from the canonical set of tensors
or any other set linked to the canonical set by a pseudo-gauge transformation. Indeed, as 
the orbital part of $\cal J$ in Eq.~(\ref{constr2}) is already taken into account in the 
energy-momentum density constraints, the only effective additional equation is
\be\label{constr3}
   n_\mu \tr \left(\wrho \, \wspt^{\mu,\lambda\nu}\right) = n_\mu \spt^{\mu,\lambda\nu}.
\ee
For the above to be actually an {\em independent} constraint, the introduction of an 
antisymmetric tensor field $\Omega_{\lambda\nu}$ is necessary. Below it is dubbed a 
{\em spin tensor potential} or shortly a {\em spin potential}~\footnote{In Ref.~\cite{flork}, 
where a hydrodynamic model of particles with spin 1/2 was proposed, this quantity 
was denoted as $\omega_{\lambda\nu}$ and dubbed the spin-polarization tensor.}. In 
analogy with the chemical potential, the components of $\Omega$ play a role of Lagrange 
multipliers coupled to the spin tensor. Including the further constraint (\ref{constr3}) 
the construction of local equilibrium density operator by means of entropy maximization
yields:
\be\label{leqd-s}
\wrho_{\rm LE} = \frac{1}{Z} \exp \left[-\int_\Sigma \di \Sigma_\mu \left(\wT^{\mu\nu} \beta_\nu 
 - \frac{1}{2} \Omega_{\lambda\nu} \wspt^{\mu,\lambda\nu} - \zeta \wj^\mu\right) \right].
\ee
One can now seek for the conditions required for the density operator to be stationary, 
i.e., for global thermodynamic equilibrium. By requiring that the integrand has zero 
divergence, one always retrieves the condition (\ref{kill}), i.e., $\beta$ ought 
to be a Killing vector and $\zeta$ to be a constant. Furthermore, if $\Omega=\varpi$, 
the form of the density operator (\ref{intrho}) depending on integral generators 
is also recovered \footnote{We note that the requirement of vanishing divergence, 
that is, stationarity in Eq.~(\ref{leqd-s}), may imply additional forms of global 
equilibrium with $\Omega \ne \varpi$. The appearance of these solutions depends on 
the symmetry features of the stress-energy and spin tensors. In particular, if the 
stress-energy tensor is symmetric, $\wT^{\mu\nu} = \wT^{\nu\mu}$, and yet the spin 
tensor does not vanish, the spin potential can be constant and independent of $\varpi$ 
or even can be non-constant, provided that it fulfills the contraction 
$(\partial_\mu \Omega_{\lambda\nu} ) \, \wspt^{\mu,\lambda\nu} = 0$. The reason for 
the appearance of such solutions is that if $\wT$ is symmetric and $\wspt \ne 0$ then 
$\partial_\mu \wspt^{\mu,\lambda\nu} = 0$, hence, there exists an additional conserved 
charge~---~the integral of the spin tensor.}.

If we now carry out a pseudo-gauge transformation with $\wPhi= \wspt$ in Eq.~(\ref{transfq}) 
to eliminate the spin tensor and recover Belinfante's stress-energy tensor (by 
Belinfante's tensors we now understand those obtained by the pseudo-gauge transformation 
with $\wPhi= \wspt$ of any original tensors $\wT^{\mu\nu} $ and $\wspt^{\mu,\lambda\nu}$)
by means of a derivation analogous to that presented in the previous section one obtains
\be\label{leqd3}
\wrho_{\rm LE} = \frac{1}{Z} \exp \left[-\int_\Sigma \di \Sigma_\mu \left(\wT^{\mu\nu}_B 
  \beta_\nu - \frac{1}{2} (\Omega_{\lambda\nu}-\varpi_{\lambda\nu})\wspt^{\mu, \lambda \nu} 
  + \frac{1}{2} \xi_{\lambda\nu} \left( \wspt^{\lambda, \mu \nu} + \wspt^{\nu, \mu \lambda}\right)  
   - \zeta \wj^\mu \right) \right],
\ee
where $\varpi$ is the thermal vorticity. Consequently, the local thermodynamic equilibrium 
operator (\ref{leqd-s}), hence (\ref{leqd3}), is equivalent to that built directly 
from Belinfante's tensor, see Eq.~(\ref{leqd}) if the following conditions are met:
\begin{enumerate}
    \item {the field $\beta$ is the same in both cases;}
    \item {the tensor $\Omega$ {\em always} coincides with thermal vorticity constructed from the $\beta$ field;}
     \item {the term involving the symmetric combination in $\lambda$ and $\nu$ indices 
     of the spin tensor vanishes.}
\end{enumerate}
The first condition is less trivial than it might look at a first glance. When imposing
the constraints (\ref{constr}) with different tensors, the field $\beta$ being a solution 
of the constraints, like e.g.
$$
  \tr (\wrho[\beta,\zeta,\ldots] \wT^{\mu\nu})n_\mu = T^{\mu\nu} n_\mu,     
$$
depends on the choice of the stress-energy tensor, namely the pseudo-gauge, and it 
is thus generally different.

The same conclusion can be reached for a more general pseudo-gauge transformation (\ref{transfq}). The
formal invariance of the operator (\ref{leqd-s}) holds if $\beta^\prime = \beta$, 
$\Omega^\prime=\Omega=\varpi$, where $\beta^\prime$ and $\Omega^\prime$ are the new
thermodynamic fields for the new set of stress-energy and spin tensors, and if the term 
proportional to $\xi$ vanishes.

Finally, we would like to point out that a density operator such as (\ref{leqd3}) involves
the appearance in the entropy current expression of a term depending on the spin 
potential and the spin tensor. Following the argument given in Ref.~\cite{betaframe} and 
assuming the extensivity of $\log Z_{\rm LE}$,
$$
  \log Z_{\rm LE} = \log \tr \left( 
  \exp \left[-\int_\Sigma \di \Sigma_\mu \left(\wT^{\mu\nu} \beta_\nu 
 - \frac{1}{2} \Omega_{\lambda\nu} \wspt^{\mu,\lambda\nu} - \zeta \wj^\mu\right) \right]
 \right) = \int_\Sigma \di \Sigma_\mu \phi^\mu,
$$
it is possible to readily obtain an entropy current $s^\mu$ from the total entropy 
$S = -\tr (\wrho_{\rm LE} \log \wrho_{\rm LE})$,
\be\label{entropycurr}
     s^\mu  = \phi^\mu + T^{\mu\nu}_{\rm LE} \beta_\nu - \zeta j^\mu_{\rm LE}
     - \frac{1}{2} \Omega_{\lambda\nu} \spt^{\mu,\lambda\nu}_{\rm LE},
\ee
where the LE subscripts stands for the mean value with the density operator 
(\ref{leqd-s}). This vector is, however, not uniquely defined as it is possible
to add vectors orthogonal to $n^\mu$ to obtain the same total entropy.
In spite of this ambiguity, the entropy current ought to be conserved when 
$T^{\mu\nu}=T^{\mu\nu}_{\rm LE}$ and $\spt^{\mu,\lambda\nu}=\spt^{\mu,\lambda\nu}_{\rm LE}$
at any time, a condition which is apparently the natural extension of the definition 
of an ideal fluid. The entropy current conservation is a consequence of the equation 
(5.8) in ref.~\cite{weert} which can be readily extended to the case with spin tensor.

\subsection{Discussion}

It is now time to pause and reflect about the physical interpretation of the discussed 
formalism. The notion of local thermodynamic equilibrium requires the existence of 
two separate space-time scales: a microscopic one over which information is not 
accessible and a macroscopic one which is used to observe system's evolution towards
global equilibrium. It is understood in the choice of the operator (\ref{leqd-s}) 
that the spin density appears among densities which ``slowly" evolve towards global 
equilibrium, just like a conserved charge density or energy density. The dissipative, 
entropy-increasing processes must drive the system to global equilibrium, hence 
(at least for systems with non-symmetric stress-energy tensor) the spin potential 
$\Omega$ should converge to the thermal vorticity (see \cite{Montenegro:2017lvf} 
for a similar discussion). Yet, this process may be slow enough so that the spin 
relaxation takes place on the same time scale as the typical dissipative hydrodynamic
process. In this case, spin density can be thought as hydrodynamically relevant, 
and the  spin potential would be a relevant hydrodynamic field. Conversely, if 
the density operator were chosen to be (\ref{leqd}), this would imply that the 
spin relaxation time is microscopically small and the value of the spin potential 
agrees with thermal vorticity.

The two situations described above can be effectively rephrased in a kinetic picture 
with colliding particles: in the first case, we would say that the spin-orbit coupling 
of the particles is much weaker than other processes responsible, for example, for 
local equilibration of their momentum. In the second case, the spin-orbit coupling 
would be as strong as any other coupling so that the spin degrees of freedom locally 
equilibrate within the same time scale as momentum. 


\begin{figure}
\includegraphics[keepaspectratio, width=0.8\columnwidth]{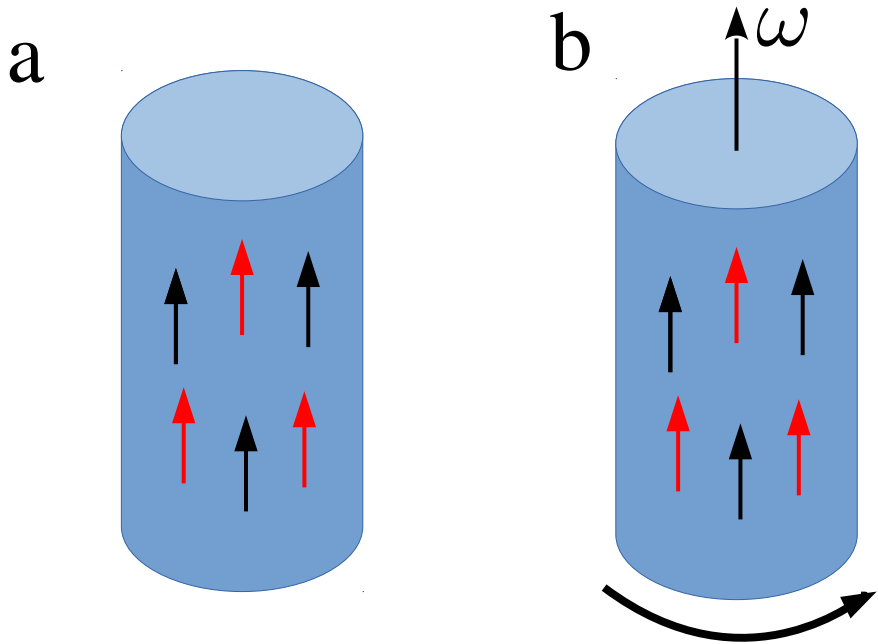}
  \caption{(Color online) a) a macroscopic fluid at rest with polarized particles 
  (black arrows) and antiparticles (red arrows). This situation, even if metastable, 
  requires a spin tensor to be described, b) if the spin tensor vanishes, the fluid 
  must be rotating (non-vanishing thermal vorticity) to have polarized particles and 
  antiparticles in the same direction.}
  \label{figura}
\end{figure}

For the purpose of illustration of the facts described above, one can envisage a fluid 
temporarily at rest with a constant temperature $T$, hence $\beta = (1/T) (1,{\bf 0})$, 
wherein both particles and antiparticles are polarized in the same direction (see Fig.~\ref{figura}). 
Such a situation cannot be described as local thermodynamic equilibrium by the density 
operator (\ref{leqd}) because the only way to have particles and antiparticles polarized 
in this case is through a non-vanishing thermal vorticity \cite{becaspin} which vanishes 
if $\beta$ is constant. In fact, if we only had Belinfante's stress-energy tensor at 
our disposal and were able to force the system to be in such an initial configuration, 
we could describe it only after a microscopic time scale, when a rotating configuration 
is established, with $\varpi \ne 0$. Conversely, with the density operator (\ref{leqd-s}) 
such a metastable situation, evolving in time, can be described by a non-vanishing spin 
potential even though thermal vorticity is vanishing.

This situation is reminiscent of the Einstein and de Haas effect: angular momentum conservation 
induces a rotation of the body because of the polarization brought about by a magnetic field. 
Yet, there is an important difference: even though the rotation occurred on a ``long" time 
scale, for the Einstein and de Haas effect to be described there is no need of a spin tensor nor 
of a spin potential, because of the absence of antimatter: the magnetization tensor $M^{\mu\nu}$ as a macroscopic density
(which, for the Dirac field, is proportional to $\ii  {\overline {\widehat \psi}} [\gamma^\mu,\gamma^\nu]  {\widehat \psi}$) 
 and the magnetic field as thermodynamic conjugate variable would 
suffice. Only when antimatter is involved, that is in a quantum relativistic field theory, 
do the spin tensor and the spin potential become necessary to describe a slow equilibration 
process of the polarization. In the limiting case of a completely neutral fluid at rest, with 
the same number of particles and antiparticles, the magnetization tensor would be zero 
and yet a metastable state with particles and antiparticles can exist. 
Mathematically, this is reflected in the linear independence of spin tensor 
(which is the dual of the axial current) and magnetization tensor in the Dirac theory.

\subsection{Relativistic hydrodynamics with spin tensor}

If we are then to describe a relativistic fluid where polarized configurations are 
not governed only by thermal vorticity, one has to introduce a spin potential tensor 
$\Omega$ among the conjugate thermodynamic fields, besides $\beta$ and $\zeta$ (the 
first steps in this direction have been done in Refs.~\cite{flork,Florkowski:2017dyn,
Florkowski:2018myy,Florkowski:2018ahw}). In the associated relativistic hydrodynamics, its 6 independent 
components have to be determined from the solutions of the 6 additional partial 
differential equations
\be\label{hydrospin}
  \partial_\lambda \spt^{\lambda,\mu\nu} = T^{\nu\mu} - T^{\mu\nu},
\ee
which appear besides the familiar ones, describing the continuity of stress-energy 
tensor and current. Of course, one needs the constitutive equations of the spin tensor, 
the stress-energy tensor, and the current in terms of $\Omega$ to be able to solve 
all hydrodynamic equations. In general, they are functional relations
\be
 j^\mu = j^\mu [\beta,\zeta,\Omega],
 \qquad T^{\mu\nu} = T^{\mu\nu} [\beta,\zeta,\Omega],
 \qquad \spt^{\lambda,\mu\nu} = \spt^{\lambda,\mu\nu} [\beta,\zeta,\Omega].
\ee
At the lowest order of approximations, the constitutive relations 
are obtained by calculating the mean values with the local equilibrium density operator 
(\ref{leqd-s}). Dissipative corrections depending on gradients of $\beta$ and $\Omega$
can be calculated with the same method outlined when discussing Eqs. (\ref{truerho})
and (\ref{dynop}).

\section{Polarization in relativistic heavy ion collisions}
\label{measur}

Does the introduction of a spin tensor in a fluid have any measurable consequence? 
This is an intriguing question with possible far-reaching consequences. Even though 
the hydrodynamic model has become one of the most useful tools for modeling of 
heavy-ion collisions, it should be stressed that neither the stress-energy tensor 
nor any other density in space and time can be directly measured or accessed. The 
actual measurements, like in any high-energy physics experiment, involve momentum 
and polarization of asymptotic particle states; the hydrodynamic model is just an 
intermediary between some initial state and the final particle spectra. In fact, 
strictly speaking, even in astrophysics one can only measure the radiation emitted 
by the plasmas and not the densities themselves. 

In fact, measurable quantities in these experiments can be generally expressed as 
expectation values of some number of creation and destruction operators of asymptotic 
states, specifically, of final hadrons in the case of relativistic heavy-ion collisions. 
For instance, the single-particle polarization matrix of a particle with momentum $p$ 
reads
\be\label{wmatrix}
 W(p)_{\sigma,\sigma^\prime} = \tr(\wrho \, a^\dagger(p)_\sigma a(p)_{\sigma^\prime}),
\ee
where $\wrho$ is the actual density operator. From the above matrix, well known quantities
such as the mean spin vector, the alignment, etc. can be calculated, including the momentum
spectrum itself by just taking the trace. Equation~(\ref{wmatrix}) makes it apparent that 
-- if any -- the dependence of measured particle spin and momenta on hydrodynamics 
with spin tensor is encoded in the density operator. 

It has been discussed in Sec.~\ref{leqdo}, that the actual density operator is the 
{\em initial local thermodynamic equilibrium} obtained from Eq.~(\ref{truerho}), 
which includes dissipative effects, as it is clear from Eq.~(\ref{dynop}). The form 
(\ref{truerho}) is hardly fit to calculate quantities like (\ref{wmatrix}) because 
field operators are to be evaluated at the initial time $\tau_0$ while the creation 
and destruction operators are those of asymptotic states. In particular, for 
relativistic heavy-ion collisions, the operators at the initial time are those of 
the quark-gluon fields while the creation and destruction operators are those of final 
hadrons. Thus, it is  necessary to make use of the form (\ref{dynop}) involving the 
effective fields at the final time $\tau$, where the effective fields are the hadronic 
ones, to carry out the calculation of Eq.~(\ref{wmatrix}). Nevertheless, the effect 
of pseudo-gauge transformations on the actual density operator, written in any of 
its equivalent forms, can be assessed by studying its effect on the initial-time 
form (the latter is more convenient to use because of its compactness). 

In a hydrodynamic approach based on Belinfante's scheme, the actual density operator
would be precisely given by (\ref{truerho}), while in a canonical-based hydrodynamics 
with relevant spin densities it would read
\be\label{truerho-s}
\wrho^{\,\prime} = \frac{1}{Z} \exp \left[-\int_{\Sigma(\tau_0)} \di \Sigma_\mu \left(\wT^{\mu\nu}_C \beta_\nu 
 - \frac{1}{2} \Omega_{\lambda\nu} \wspt^{\mu,\lambda\nu}_C - \zeta \wj^\mu\right) \right].
\ee
Equation~(\ref{truerho-s}) can be transformed in the same way as in the previous 
section and turned into an expression like Eq.~(\ref{leqd3}), namely
\be\label{truerho-s2}
\wrho^{\,\prime} = \frac{1}{Z} \exp \left[-\int_{\Sigma(\tau_0)} \di \Sigma_\mu \left(\wT^{\mu\nu}_B 
  \beta_\nu - \frac{1}{2} (\Omega_{\lambda\nu}-\varpi_{\lambda\nu})\wspt^{\mu, \lambda \nu}_C 
  + \frac{1}{2} \xi_{\lambda\nu} \left( \wspt^{\lambda, \mu \nu}_C + \wspt^{\nu, \mu \lambda}_C\right)  
   - \zeta \wj^\mu \right) \right].
\ee
Therefore, for any measurable quantity $\widehat X$, there would be a difference 
between the mean values calculated with the density operators (\ref{truerho}) and (\ref{truerho-s2}). Defining:
\be
  \widehat{A} = \int_{\Sigma(\tau_0)} \di \Sigma_\mu \left(\wT^{\mu\nu}_B 
  \beta_\nu  - \zeta \wj^\mu\right) \\
\ee
and
\be
 \widehat{B} = \int_{\Sigma(\tau_0)}  \di \Sigma_\mu \left[- \frac{1}{2} 
 (\Omega_{\lambda\nu}-\varpi_{\lambda\nu})\wspt^{\mu, \lambda \nu}_C 
  + \frac{1}{2} \xi_{\lambda\nu} \left( \wspt^{\lambda, \mu \nu}_C + 
  \wspt^{\nu, \mu \lambda}_C\right) \right] ,
\ee
one has
\be\label{deltax}
 \Delta X = \tr(\wrho^{\,\prime} \widehat{X}) - \tr (\wrho \,\widehat{X}) = 
 \frac{1}{Z'}\tr(\e^{\widehat{A}+\widehat{B}} \widehat{X}) - \frac{1}{Z} 
  \tr (\e^{\widehat{A}} \widehat{X}) \simeq 
  \int_0^1 \di z \; \tr \left[ \wrho \, \widehat{X} \e^{z \widehat A} 
 \left( \widehat{B} - \tr(\wrho \,\widehat{B}) \e^{-z\widehat{A}} \right) \right],
\ee
where the right-hand side is the leading term in the linear response theory.
If $\wspt_C$ is the canonical spin tensor of the Dirac field, the term in $\xi$
vanishes because $\wspt_C$ is a completely antisymmetric tensor and, hence, the difference 
in the theoretical value depends on the correlator between $\widehat{X}$ and $\widehat{B}$, 
that is the integral of the spin tensor weighted by the difference between spin potential 
and thermal vorticity. The evaluation of (\ref{deltax}) is not an easy task but it can be 
envisaged that this difference will be a quantum effect and a tiny one for most observables 
which mostly depend on the $\beta$ field, namely velocity and temperature field, such as the 
momentum spectra. In fact, the difference could be significant for observables which depend 
linearly on the thermal vorticity, such as polarization.

\section{Conclusions}
\label{discu}

In summary, we have shown that the non-equilibrium or local-equilibrium thermodynamics is 
sensitive to the pseudo-gauge transformation of the stress-energy and spin tensors in quantum 
field theory. This conclusion is a considerable extension of previous arguments 
\cite{becatinti1,becatinti2}, as we have shown here that pseudo-gauge transformations quantitatively 
affect theoretical values of measurable quantities in high-energy physics experiments, especially 
polarization of final particles created in relativistic heavy-ion collisions. 

The inequivalence of different stress-energy tensors arises most clearly in the description 
of a completely neutral fluid at rest with finite polarization of both particles and 
antiparticles --- this corresponds to a metastable local equilibrium state that has a finite 
value of some spin tensor and zero (thermal) vorticity. If the stress-energy tensor used is 
Belinfante's symmetrized one, finite polarization should always be accompanied by a macroscopic 
rotation. 

This psuedo-gauge inequivalence has a hydrodynamic counterpart. Hydrodynamics of metastable 
polarized neutral fluids requires a rank two spin potential tensor $\Omega_{\mu\nu}$ as conjugate 
thermodynamic field to the spin tensor. This addition of course complicates the standard 
hydrodynamic equations to be solved. 

Finally, it should be pointed out that the question remains how to determine the 
''right" spin tensor among all possible ones. Indeed, the canonical spin tensor obtained 
from Eq.~(\ref{SC}) is one particular option; other spin tensors can be obtained by 
applying a pseudo-gauge transformation and they will not yield the same local thermodynamic 
equilibrium operator, according to the discussion in Sect.~\ref{ineq} following 
Eq.~(\ref{leqd3}). Apparently, the issue cannot be settled theoretically, but 
experimentally comparing measured quantities to the different predictions. We note 
that similar issues have been very intensively studied in the context of the proton 
spin decomposition, for example see Refs.~\cite{Chen:2008ag,Leader:2013jra}.

\section*{Acknowledgments}

Very useful discussions with  B. Friman, M. Stephanov, L. Tinti are gratefully acknowledged. 
W.F. was supported in part by the Polish National Science Center Grant No. 2016/23/B/ST2/00717.
E.S.  was  supported  by  BMBF  Verbundprojekt 05P2015 - Alice at High Rate and by the Deutsche 
Forschungsgemeinschaft  (DFG)  through  the  grant  CRC-TR 211 ``Strong-interaction matter 
under extreme conditions'' and by the COST Action CA15213 “Theory of hot matter and 
relativistic heavy-ion collisions”  (THOR).




\appendix


\begin{thebibliography}{99}
\section*{References}

\bibitem{STAR}
  L.~Adamczyk {\it et al.} [STAR Collaboration],
  Nature {\bf 548} (2017) 62.

\bibitem{becacsernai}
  F.~Becattini, L.~Csernai and D.~J.~Wang,
  Phys.\ Rev.\ C {\bf 88} (2013) no.3,  034905
   Erratum: [Phys.\ Rev.\ C {\bf 93} (2016) no.6,  069901].

\bibitem{becavort}
  F.~Becattini {\it et al.},
  Eur.\ Phys.\ J.\ C {\bf 75} (2015) no.9,  406
  Erratum: [Eur.\ Phys.\ J.\ C {\bf 78} (2018) no.5,  354].

\bibitem{Pang:2016igs}
  L.~G.~Pang, H.~Petersen, Q.~Wang and X.~N.~Wang,
  Phys.\ Rev.\ Lett.\  {\bf 117} (2016) no.19,  192301.

\bibitem{Deng:2016gyh}
  W.~T.~Deng and X.~G.~Huang,
  Phys.\ Rev.\ C {\bf 93} (2016) no.6,  064907.
  
\bibitem{Karpenko:2016jyx}
  I.~Karpenko and F.~Becattini,
  Eur.\ Phys.\ J.\ C {\bf 77} (2017) no.4,  213.

\bibitem{Xie:2017upb}
  Y.~Xie, D.~Wang and L.~P.~Csernai,
  Phys.\ Rev.\ C {\bf 95} (2017) no.3,  031901.

\bibitem{Li:2017slc}
  H.~Li, L.~G.~Pang, Q.~Wang and X.~L.~Xia,
  Phys.\ Rev.\ C {\bf 96} (2017) no.5,  054908.

\bibitem{Sun:2017xhx}
  Y.~Sun and C.~M.~Ko,
  Phys.\ Rev.\ C {\bf 96} (2017) no.2,  024906.

\bibitem{Lan:2017nye}
  S.~Lan, Z.~W.~Lin, S.~Shi and X.~Sun,
  Phys.\ Lett.\ B {\bf 780} (2018) 319.

\bibitem{Xia:2018tes}
  X.~L.~Xia, H.~Li, Z.~B.~Tang and Q.~Wang,
  Phys.\ Rev.\ C {\bf 98} (2018) 024905.

\bibitem{becaspin}
   F.~Becattini, V.~Chandra, L.~Del Zanna and E.~Grossi,
  Annals Phys.\  {\bf 338} (2013) 32.

\bibitem{STAR2}
  J.~Adam {\it et al.} [STAR Collaboration],
  Phys.\ Rev.\ C {\bf 98} (2018) 014910.

\bibitem{QMstar}
   T.~Niida [STAR Collaboration],
  arXiv:1808.10482 [nucl-ex].
  
\bibitem{flork}
  W.~Florkowski, B.~Friman, A.~Jaiswal and E.~Speranza,
   Phys.\ Rev.\ C {\bf 97} (2018) no.4,  041901.
   
\bibitem{hehl}
  F.~W.~Hehl, Rept.\ Math.\ Phys.\  {\bf 9}, 55 (1976).   
   
\bibitem{becatinti1}   
  F.~Becattini and L.~Tinti,
  Phys.\ Rev.\ D {\bf 84} (2011) 025013.
   
\bibitem{becatinti2}   
  F.~Becattini and L.~Tinti,
  Phys.\ Rev.\ D {\bf 87} (2013) no.2,  025029.

\bibitem{zubarev}
  D.~N.~Zubarev, Sov. Phys. Doklady {\bf 10}, 850 (1966);
  D.~N.~Zubarev, A.~V.~Prozorkevich, S.~A.~Smolyanskii, Theoret. and Math. 
  Phys. 40 (1979), 821.

\bibitem{weert}
 Ch.~G.~Van~Weert, Ann.\ Phys.\ {\bf 140}, 133 (1982).
 
\bibitem{betaframe}
  F.~Becattini, L.~Bucciantini, E.~Grossi and L.~Tinti,
  Eur.\ Phys.\ J.\ C {\bf 75} (2015) no.5,  191.
  
\bibitem{hongo}  
 M.~Hongo,
   Annals Phys.\  {\bf 383} (2017) 1.

 \bibitem{becacov}
   F.~Becattini,
   Phys.\ Rev.\ Lett.\  {\bf 108} (2012) 244502.
 
\bibitem{hosoya}
  A.~Hosoya, M.~Sakagami and M.~Takao, Annals Phys.\  {\bf 154}, 229 (1984). 

\bibitem{lorce}
  C.~Lorc\'e, Eur.\ Phys.\ J.\ C {\bf 78} (2018) no.9,  785

\bibitem{Montenegro:2017lvf}
  D.~Montenegro, L.~Tinti and G.~Torrieri,
  Phys.\ Rev.\ D {\bf 96} (2017) 076016.
   
\bibitem{Florkowski:2017dyn} 
  W.~Florkowski, B.~Friman, A.~Jaiswal, R.~Ryblewski and E.~Speranza, Phys.\ Rev.\ D {\bf 97} (2018) 116017.
  
\bibitem{Florkowski:2018myy}
  W.~Florkowski, E.~Speranza and F.~Becattini,
  Acta Phys.\ Polon.\ B {\bf 49} (2018) 1409.
  
\bibitem{Florkowski:2018ahw} 
  W.~Florkowski, A.~Kumar and R.~Ryblewski,
  Phys.\ Rev.\ C {\bf 98} (2018) 044906.
  
\bibitem{Chen:2008ag}
  X.~S.~Chen, X.~F.~Lu, W.~M.~Sun, F.~Wang and T.~Goldman,
  Phys.\ Rev.\ Lett.\  {\bf 100} (2008) 232002.
 
\bibitem{Leader:2013jra} 
  E.~Leader and C.~Lorce,
  Phys.\ Rept.\  {\bf 541}  (2014) 163.
  

   
\end{thebibliography}
\end{document}